\documentclass[sigconf]{acmart}
\usepackage{bigstrut}
\usepackage{booktabs}
\usepackage{multirow}
\usepackage{microtype}
\usepackage{graphicx}
\usepackage{subfig}
\usepackage{overpic}
\usepackage{balance}
\usepackage{amsthm}
\usepackage{array}
\usepackage{clrscode}
\usepackage{multicol}
\usepackage{float}
\usepackage{newfloat}
\usepackage{color}
\usepackage{xcolor}
\usepackage{amsopn}
\usepackage{mathrsfs}
\usepackage{mathtools}
\usepackage{amsmath}
\usepackage{arydshln}
\usepackage{blkarray}
\usepackage{enumerate}
\usepackage{courier}
\usepackage{rotating}
\usepackage{bm}
\usepackage{wrapfig}
\usepackage{ragged2e}
\usepackage[misc]{ifsym}
\usepackage{threeparttable}
\usepackage{makecell}
\usepackage{adjustbox} %
\usepackage{enumitem}
\usepackage{hyperref}
\usepackage{url}
\usepackage{xspace}
\usepackage{comment}

\usepackage{algorithm}
\usepackage{algpseudocode} 
\usepackage{algorithmicx} 
\usepackage[ruled, vlined, nofillcomment, linesnumbered, algo2e]{algorithm2e}

\definecolor{codegreen}{rgb}{0,0.3,0.6}
\definecolor{codegray}{rgb}{0.5,0.5,0.5}

\newcommand{\ie}{\emph{i.e.,}\xspace}
\newcommand{\eg}{\emph{e.g.,}\xspace}

\newcommand{\paratitle}[1]{\vspace{1.5ex}\noindent\textbf{#1}}

\newcommand{\ignore}[1]{}

\AtBeginDocument{%
  \providecommand\BibTeX{{%
    \normalfont B\kern-0.5em{\scshape i\kern-0.25em b}\kern-0.8em\TeX}}}

\setcopyright{acmcopyright}
\copyrightyear{2023}
\acmYear{2023}
\acmDOI{XXXXXXX.XXXXXXX}
\acmConference[RecSys '23]{Proceedings of the 17th ACM Conference on Recommender Systems}{September 18--23, 2023}{Singapore}
\acmBooktitle{Proceedings of the 17th ACM Conference on Recommender Systems (RecSys '23), September 18--23, 2023, Singapore}
\acmPrice{15.00}
\acmISBN{978-1-4503-XXXX-0/22/xx}

\begin{document}

\title{Reciprocal Sequential Recommendation}

\author{Bowen Zheng$^{\dagger}$}
\email{bwzheng0324@gmail.com}
\affiliation{%
    \institution{Gaoling School of Artificial Intelligence, Renmin University of China}
    \city{Beijing}
    \country{China}
}

\author{Yupeng Hou$^{\dagger}$}
\email{houyupeng@ruc.edu.cn}
\affiliation{%
    \institution{Gaoling School of Artificial Intelligence, Renmin University of China}
    \city{Beijing}
    \country{China}
}

\author{Wayne Xin Zhao$^{\dagger\ddagger}$
\textsuperscript{\Letter}
}
\email{batmanfly@gmail.com}
\affiliation{
    \institution{Gaoling School of Artificial Intelligence, Renmin University of China}
    \city{Beijing}
    \country{China}
}

\author{Yang Song \textsuperscript{\Letter}}
\email{songyang@kanzhun.com}
\affiliation{%
    \institution{BOSS Zhipin}
    \city{Beijing}
    \country{China}
}

\author{Hengshu Zhu}
\email{zhuhengshu@kanzhun.com}
\affiliation{%
    \institution{BOSS Zhipin}
    \city{Beijing}
    \country{China}
}

\thanks{$\dagger$ Beijing Key Laboratory of Big Data Management and Analysis Methods.}
\thanks{$\ddagger$ Beijing Academy of Artificial Intelligence, Beijing, 100084, China.}
\thanks{\Letter\ Corresponding authors.}

\renewcommand{\authors}{Bowen Zheng, Yupeng Hou, Wayne Xin Zhao,   Yang Song, Hengshu Zhu}
\renewcommand{\shortauthors}{Bowen Zheng et al.}

\begin{abstract}
Reciprocal recommender system (RRS), considering a two-way matching between two parties, has been widely applied in  online platforms like online dating and recruitment.  
Existing RRS models mainly 
capture static user preferences, which have neglected the evolving user tastes and the dynamic matching relation between the two parties. 
 Although dynamic user modeling has been well-studied in sequential recommender systems, existing solutions are developed in a user-oriented manner. Therefore, it is non-trivial to adapt sequential recommendation algorithms to reciprocal recommendation. 
In this paper, we formulate  RRS as a distinctive sequence matching task, and further propose a new approach \emph{ReSeq} for RRS, which is short for \underline{Re}ciprocal \underline{Seq}uential recommendation.
To capture dual-perspective matching, 
we propose to learn fine-grained sequence similarities by co-attention mechanism \textcolor{black}{across different time steps.} 
Further, to improve the inference efficiency, we introduce the self-distillation technique \textcolor{black}{to distill knowledge from the fine-grained matching module into the more efficient student module. In the deployment stage, only the efficient student module is used, greatly speeding up the similarity computation.} 
Extensive experiments on five real-world datasets from two scenarios demonstrate the effectiveness and efficiency of the proposed method.
Our code is available at \textcolor{blue}{\url{https://github.com/RUCAIBox/ReSeq/}}.
\end{abstract}



\maketitle

\section{Introduction}

Reciprocal Recommender System (RRS)~\cite{pizzato2010recon} has been widely deployed in online platforms, such as online dating~\cite{pizzato2010recon,xia2015reciprocal,neve2019latent} and recruitment~\cite{le2019towards,jiang2020learning,yang2022modeling}.
Different from traditional recommendation algorithms, it emphasizes a two-way (or bilateral) selection process when establishing the matching between the two parties involved in recommender systems. 
Generally, existing RRS works can be roughly divided into content-based methods and collaborative filtering-based methods. 
The former mainly model user interests from their
personalized profiles~\cite{pizzato2010recon,luo2020rrcn,yildirim2021bideepfm}, while the latter usually learn static user representations to capture long-term user preference based on collaborative filtering~\cite{xia2015reciprocal,neve2019latent,luo2020motif}. 
However, the interests of users tend to dynamically shift along with alterations in the surrounding environment over time, which is disregarded by the above two categories of models.  
For instance, in the context of online recruitment, as individuals gain more work experience and qualifications, they are inclined to increase the criteria in job hunting. 
Therefore, it is essential to effectively model the dynamic preference changes in such a two-way preference selection process, thus making more accurate recommendations that can satisfy the needs of both parties.

To model the dynamic interests of users, sequential recommendation methods mainly capture user preference by modeling the   chronological user behavior sequences~\cite{kang2018self,sun2019bert4rec} in a unilateral setting. 
As a typical approach, they utilize the output in the last time step of the sequential recommendation model for matching prediction. 
Different from conventional recommendation algorithms, which are mainly developed based on the single-perspective active selection from a user to items, 
users play  a dual role in reciprocal recommendation, serving not only as the party to actively select but also as the other party to be passively selected.  
Thus, the RRS necessitates the dynamic modeling of both parties, especially in the context of two-way matching, which is likely to result in a more difficult learning problem.

As the technical approach, we formulate reciprocal recommendation as a distinctive sequence matching task, where the recommendation prediction is performed based on the matching between the behavior sequences of the involved two parties. 
Specifically, we establish the behavior sequences in dual perspectives, 
and then conduct the two-sided matching  by modeling fine-grained sequential semantic interactions.
While, it takes a high time cost to model fine-grained matching, and we further consider employing self-distillation techniques (commonly used in model compression~\cite{zhang2019your,xu2020improving,liu2020fastbert}) to improve model efficiency.

Inspired by the above motivations, in this paper, we propose \emph{ReSeq}, a \underline{Re}ciprocal \underline{Seq}uential recommendation method.
There are three major technical contributions in our approach.
Firstly, we model bilateral behavior sequences by a transformer network\cite{vaswani2017attention} based on user's  dual-perspective embeddings (\ie \emph{active} and \emph{passive}).
 Secondly, we design a multi-scale matching prediction method to capture fine-grained sequential interactions at both macro and micro levels. 
Finally,  we largely improve the efficiency of the proposed model via micro-to-macro self-distillation,  while ensuring accuracy and effectiveness.

To the best of our knowledge, this is the first time that dynamic behavior sequences of two parties  are considered in reciprocal recommendation, which can achieve both accurate and efficient recommendations. Extensive experiments on five datasets from two reciprocal recommendation scenarios demonstrate the effectiveness of our proposed method.

\section{Related Work}
In this section, we review the related work in three aspects, namely Reciprocal Recommendation, Sequential Recommendation and Self-Distillation.

\subsection{Reciprocal Recommendation}
Reciprocal recommender system (RRS)~\cite{pizzato2010recon} is a recommender system between people and people that significantly deviates from conventional user-item recommender systems. 
To emphasize the mutual matching relationship between users, RRS requires modeling the interests of both parties in the recommendation, which poses a greater challenge for delivering effective recommendations.
Existing RRS studies can be roughly categorized into two groups: content-based methods~\cite{pizzato2010recon,alanazi2013people,tu2014online} and collaborative filtering-based methods~\cite{cai2012reciprocal,xia2015reciprocal}. Content-based methods focus on the matching of user profiles,
while collaborative filtering-based methods pay more attention to modeling historical user interactions. 
There also exist hybrid models that integrate user behavior modeling as well as content matching~\cite{akehurst2011ccr,yang2017combining,kleinerman2018optimally,hou2022leveraging}. 
In recent years, deep learning techniques have been increasingly used to improve the expressive power of RRS models, including matrix factorization~\cite{qu2018reciprocal}, latent factor models~\cite{neve2019latent}, factorization machines~\cite{yildirim2021bideepfm}, and graph neural networks (GNNs)~\cite{luo2020motif,yang2022modeling} 
Furthermore, several studies also make use of multiple types of user behaviors to learn better bilateral user preferences in RRS scenarios~\cite{le2019towards,fu2021beyond}. 
Although effective to some extent,
existing methods mainly encode user behaviors and profiles into static representations to model long-term user interests, ignoring the dynamic changes over time. In this work, we formulate reciprocal recommendation as a distinctive sequence matching task and propose to leverage bilateral behavior sequences for dynamic interest modeling on both sides.

\subsection{Sequential Recommendation} \label{sec:seqrec}
In contrast to conventional collaborative filtering algorithms based on all historical user-item interactions~\cite{koren2008factorization,kabbur2013fism,he2018nais,rendle2009bpr,he2017neural,he2020lightgcn}, sequential recommendation pays more attention to the temporal aspects of user behaviors~\cite{hidasi2015session,sun2019bert4rec,kang2018self}.
Early sequential recommendation methods are frequently founded upon Markov Chains techniques~\cite{rendle2010factorizing,he2016fusing}. With the fast development of deep neural networks, numerous deep models like Recurrent Neural Networks (RNNs) are employed to facilitate item sequence modeling~\cite{hidasi2015session,tang2018personalized,yuan2019simple}. 
More recently, GNN-based~\cite{wu2019session,xu2019graph} and Transformer-based methods~\cite{kang2018self,sun2019bert4rec,hou2022core} also show great effects on sequential recommendation task. 
Besides direct item sequence modeling, recent studies also exploited techniques to additionally enhance model effectiveness and robustness, such as pre-training~\cite{zhou2020s3,hou2022towards}, data augmentation~\cite{wang2021counterfactual} and regularized training~\cite{xie2022cl4rec,zhou2023mvs}.

The above sequential recommendation methods
have achieved amazing performance in conventional user-item recommendation. 
Nevertheless, in the reciprocal recommendation scenario, only single-perspective modeling usually fails to achieve satisfactory results. 
The other possible reason is that compared with the static item target in user-item recommendation, the dynamic characteristics of both users in reciprocal recommendation require more detailed interactions between both parties.
In our approach, we model user behavior sequences from both active and passive perspectives, and a multi-scale matching prediction technique is proposed for comprehensive and fine-grained two-way interactions.

\subsection{Self-Distillation}
Knowledge distillation~\cite{hinton2015distilling} is a popular technique that can effectively transfer knowledge from complex teacher networks to more streamlined student networks. In the realm of recommender systems, it is often used for domain knowledge transfer~\cite{yang2022cross,tu2021conditional} or model compression~\cite{kang2021topology,tian2023directed}. However, such methods mainly necessitate meticulously crafted teacher models, which incur additional design and experimental costs. 
In recent years, the self-distillation technique~\cite{zhang2019your} that distills knowledge from itself has been proposed, and there are three prevalent self-distillation methods exist: \textbf{(a)} distillation from deep to shallow layers of the same network~\cite{zhang2019your,luan2019msd,liu2020fastbert}; \textbf{(b)} using the self-ensemble model (from the most recent time steps) as a virtual teacher model~\cite{xu2020improving,shen2022self}; \textbf{(c)} data augmentation-based methods~\cite{huang2020comprehensive,xu2019data}.
In comparison to conventional knowledge distillation, self-distillation remains an area that is worth exploring within the realm of recommendation systems~\cite{xia2022device}.
In order to ensure low time consumption while engaging in more fine-grained two-way interactions, we put forth a simple self-distillation framework in this paper.
Our approach is most similar to the first of the above self-distillation methods, but instead of using the deeper network outputs as knowledge, we use time-sensitive fine-grained micro-level matching as a teacher for efficient macro-level matching.

\begin{figure*}[t!]
\centering
\includegraphics[width=0.91\linewidth]{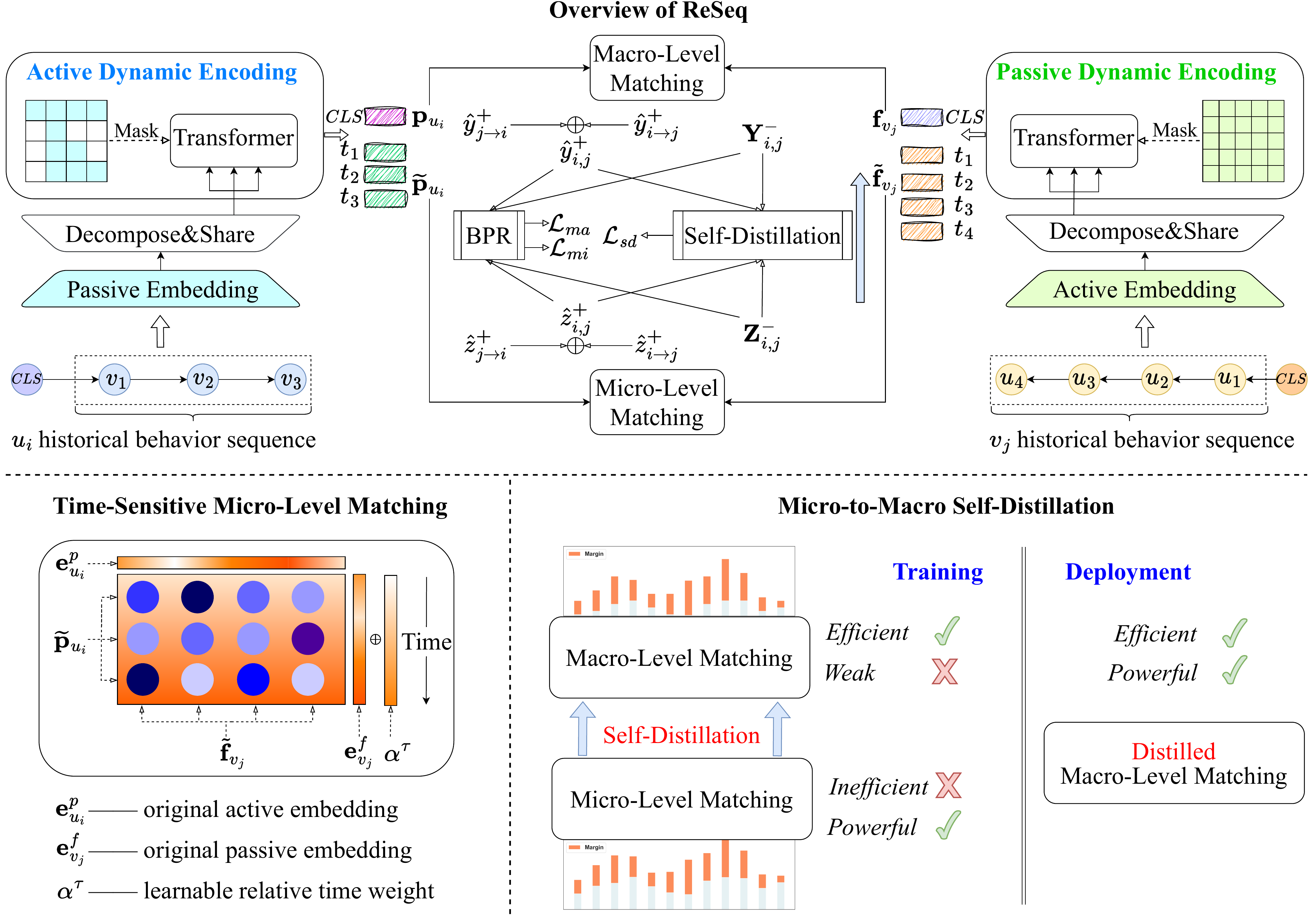}
\caption{The overall framework of our ReSeq. Note that it only shows the single perspective of the model from $u_i$ to $v_j$, the other perspective from $v_j$ to $u_i$ is symmetrical.}
\label{fig:model}
\end{figure*}

\section{Methodology}
In this section, we first formulate our reciprocal sequential recommendation task, and secondly introduce the proposed ReSeq method in detail. The overall architecture is shown in Figure~\ref{fig:model}.

\subsection{Problem Formulation} \label{sec:pb_fmu}
In conventional recommender systems, the goal of recommendation algorithms is typically to predict the user's interest in items (movies, books, clothes, etc), which essentially refers to user-oriented item recommendation. 
Such a recommendation paradigm is solely based on the user perspective for item selection.  
As a comparison, Reciprocal Recommender System (RRS)~\cite{pizzato2010recon} targets a two-sided recommendation setting, in which either side can select the users or resources on the other side, 
\eg online dating recommendation between  females and males and job recommendation between job seekers and recruiters. 
Formally, we denote the two parties in reciprocal recommendation as $u_i \in \mathcal{U}$ (the first side)  and $v_j \in \mathcal{V}$ (the second side), respectively.
In this setting, we do not explicitly use the term of \emph{item}, since each side can play the roles of \emph{users} (those who actively select the interactions) and \emph{items} (those who are interacted with passively) as in traditional recommender systems. 
Instead, we call both sides \emph{users} and use different notations to discriminate the two sides. 
A successful recommendation in RRS indicates that: \textbf{(a)} the user $u_i$ demonstrates interest in the user $v_j$; \textbf{(b)} the features of $u_i$ align with the preferences and requirements of $v_j$. 
This process can be characterized as a dual-perspective active selection process~\cite{pizzato2010recon,xia2015reciprocal,neve2019latent,yang2022modeling}, the goal of RRS typically aims at learning a joint function $\hat{y}_{i,j}= \mathop{g}\limits^{\rightarrow}(u_i, v_j) +  \mathop{g}\limits^{\leftarrow}(u_i, v_j) $ that can  model this bilateral preference selection for the recommendation. 
In this paper, our approach is established in the aforementioned setting for RRS. 
This formulation characterizes a dual-perspective sequence matching task for the RRS, which is distinctive from previous settings~\cite{pizzato2010recon}.

To be more specific, at every time step, each user possesses a historical behavior sequence up to the current time $T$, denoted by $\mathcal{S}^T_{u_i} = (v^{t_1}_1, v^{t_2}_2,..., v^{t_a}_a)$ or $\mathcal{S}^T_{v_j} = (u^{t_1}_1, u^{t_2}_2,..., u^{t_b}_b)$ where $t_a < T$ and $t_b < T$ are the timestamps for interactions. In the sequence $\mathcal{S}^T_{u_i}$, $v^{t_{j'}}_{j'}$ represents that $u_i$ is matched with $v_{j'}$ at time $t_{j'}$. And vice versa, $u^{t_{i'}}_{i'}$ in $\mathcal{S}^T_{v_j}$ represents the historical matching relationship with $v_j$.
Our target is to construct a matching function $\hat{y}_{i,j}=\mathop{g}\limits^{\rightarrow}(u_i, {S}^T_{u_i}, v_j, {S}^T_{v_j}) + \mathop{g}\limits^{\leftarrow}(u_i, {S}^T_{u_i}, v_j, {S}^T_{v_j})$ that predicts the matching degree or score $\hat{y}_{i,j}$ between both users by dual-perspective modeling of their historical behavior sequences.

\subsection{Dual-Perspective Dynamic Sequential Behavior Modeling} 
\label{sec:bhv_model}

Existing RRS methods primarily utilize user profiles for prediction matching~\cite{pizzato2010recon,luo2020rrcn} or learning user static representations based on historical user interactions~\cite{xia2015reciprocal,neve2019latent}.
However, such methods tend to mainly focus on capturing long-term user interests, thereby ignoring the dynamic changes in user preferences over time.
Our approach models reciprocal recommendation as a distinctive sequence matching task, where the matching of both sides is conducted based on their dynamic behavior sequences.
In this section, we present our method for modeling user behavior sequences under dual perspectives.
To be more precise, we first illustrate the user dual-perspective representations in reciprocal recommendation, and then we elucidate the process of encoding both active and passive aspects of user behavior sequences.

\subsubsection{Bilateral User Representation Learning}
As mentioned in the previous section, the matching of users on both sides in reciprocal recommendation is a dual-perspective process.
This phenomenon results in the user playing a dual role in reciprocal recommendation, serving not only as the active selecting party but also as the passive accepting party.
Hence, methods in reciprocal recommendation frequently use dual-perspective user representations~\cite{neve2019latent,yang2022modeling}.
 
\paratitle{User representation learning under dual perspectives.}
For both users in the reciprocal recommendation scenario, there exist two distinct representations: one is when they act as active selectors by displaying their preferences over the other party, and the other is when they act as passive candidates by exhibiting their own features for being selected.
For simplicity, we use ``\emph{\underline{p}reference}'' as the marker for the user active representations and ``\emph{\underline{f}eature}'' as the marker for the user passive representations.
Particularly, for users $\mathcal{U}$ on one side, we maintain two separate embedding matrices $\bm{M}^{U}_{p} \in \mathbb{R}^{|\mathcal{U}| \times d}$ and $\bm{M}^{U}_{f} \in \mathbb{R}^{|\mathcal{U}| \times d}$, which represent user active and passive embeddings, respectively. Likewise, there are also two embedding matrices $\bm{M}^{V}_{p} \in \mathbb{R}^{|\mathcal{V}| \times d}$ and $\bm{M}^{V}_{f} \in \mathbb{R}^{|\mathcal{V}| \times d}$ for the users $\mathcal{V}$ on the other side.

\paratitle{Decomposed and shared bilateral user embedding.}
As discussed previously, the matching of both parties in reciprocal recommendation is a dual-perspective alignment of their active interests and passive attractions.
This implies that the representation space of bilateral users should be properly aligned.
Formally, the above embedding matrices are aligned between $\bm{M}^{U}_{p}$ and $\bm{M}^{V}_{f}$, as well as between $\bm{M}^{V}_{p}$ and $\bm{M}^{U}_{f}$.
To reflect this dual-perspective alignment relationship, we perform a decomposition of the user embedding matrix, and part of the decomposed matrix is shared between the dual perspectives. Take a perspective as an example:
\begin{align}
    \bm{M}^{U}_{p} = \bm{A} \cdot \bm{C}, \ \ \ \ \
    \bm{M}^{V}_{f} = \bm{B} \cdot \bm{C},
    \label{eq:decomp}
\end{align}
where $\bm{A} \in \mathbb{R}^{|\mathcal{U}| \times {d'} }$, $\bm{B} \in \mathbb{R}^{|\mathcal{V}| \times {d'}}$, and $\bm{C} \in \mathbb{R}^{{d'} \times d}$ is shared between the user active representations on one side and the user passive representations on the other side. Another perspective also applies the identical approach of matrix decomposition and sharing.

\subsubsection{Dual-Perspective Behavior Sequence Encoding}
\label{sec:seq_encode}
After obtaining the decomposed and shared dual-perspective user embedding, we next encode the user behavior sequence by proposing two different methods. The purpose is to actively and passively model behavior sequences according to the different dynamic characteristics of the two roles for the user.

\paratitle{Unidirectional active dynamic encoding.}
When a user assumes the role of the active party, each historical matching user in the behavior sequence acts as the passive party.
Intuitively, the features and characteristics of passive candidates in recommendations are indicative of user interests and preferences.
Therefore, we encode the active dynamic representations of the user based on the passive representations of historical matching users in the behavior sequence.

Given a behavior sequence $\mathcal{S}^T_{u_i} = (v^{t_1}_1, v^{t_2}_2,..., v^{t_a}_a)$, we first pad the sequence to a fixed length $n$, and then add a special token ($[CLS]$) at the beginning of the sequence to aggregate the sequence representations. 
Next, we apply the lookup operation from $\bm{M}^{V}_{f}$ along with a learnable position embedding $P \in \mathbb{R}^{(n+1) \times d}$ to form the input sequence passive embeddings $\bm{E}^{V}_{f} \in \mathbb{R}^{(n+1) \times d}$. 
Subsequently, we exploit a widely-used Transformer network~\cite{vaswani2017attention} for sequence modeling. Concretely, a typical Transformer architecture consists of multiple layers of masked multi-head self-attention (denoted by Masked-MHAttn(·)), and each layer of self-attention followed by a point-wise feed-forward network (denoted by FFN(·)). The above sequence embeddings $\bm{E}^{V}_{f}$ combined with position encoding are used as the initial input of the Transformer. Then, the calculations can be formally written as:
\begin{align}
    \bm{H}^{l} & =\operatorname{FFN}\left(\operatorname{Masked-MHAttn}\left(\bm{H}^{l-1}, \bm{W}_{mask}\right)\right), \ l \in\{1, \ldots, L\},
\end{align}
where $\bm{H}^{l}$ is the hidden states in the $l$-th layer, and $\bm{H}^{0} = \bm{E}^{V}_{f}$. $\bm{W}_{mask} \in \mathbb{R}^{(n+1) \times (n+1)}$ is the attention mask weight matrix used to regulate the attention object for each part within the sequence. 

In our case, to encode the time-sensitive active dynamic behaviors, we use a unidirectional attention mask weight matrix which allows the model to concentrate more on recent user behaviors. 
As shown in Figure~\ref{fig:model}, compared with the diagonal matrix unidirectional mask, our mask matrix separately processes the corresponding row and column of the $[CLS]$ token to ensure both the order of information transmission and the global aggregation of $[CLS]$ representation. 
Note that the mask matrix used in our approach is equivalent to adding the $[CLS]$ token at the end of the sequence. However, compared to individually inserting $[CLS]$ token to behavior sequences of different lengths, our approach is more convenient for input construction and model implementation.

Ultimately, the outputs from the final layer of the $L$-layer Transformer network serve as the active dynamic representations of the user. Additionally, we divide $\bm{H}^{L}$ into two scales:
\begin{align}
    \bm{p}_{u_i} = \bm{H}^{L}_{0}, \ \ \ \ \
    \widetilde{\bm{p}}_{u_i} = \bm{H}^{L}_{1:n+1},
\end{align}
where $\bm{H}^{L}_{0}$ is the corresponding output of $[CLS]$ token, $\bm{p}_{u_i} \in \mathbb{R}^{d}$ indicates the user's macro or global active dynamic representation. And $\bm{H}^{L}_{1:n+1}$ is the other part of the outputs, $\widetilde{\bm{p}}_{u_i} \in \mathbb{R}^{n \times d}$ represents the user's micro or fine-grained active dynamic representations.

\paratitle{Bidirectional passive dynamic encoding.}
Similar to the unidirectional active dynamic encoding described above, we also employ a Transformer network to model the passive perspective of user behavior sequences.
The difference is that, in this context, each historical matching user in the behavior sequence assumes the role of the active party.
Empirically, a successful matching means that the features and characteristics of the current user meet the preferences and requirements of the active selector.
We utilize the lookup operation from $\bm{M}^{V}_{p}$ instead of $\bm{M}^{V}_{f}$ to form the input sequence active embeddings $\bm{E}^{V}_{p} \in \mathbb{R}^{(n+1) \times d}$.

Furthermore, in contrast to the time-sensitivity of user active dynamic behaviors, when a user acts as a passive individual, his features and characteristics tend to remain relatively stable over time.
Thus, as shown on the right side of Figure~\ref{fig:model}, we encode the passive dynamic representations by a bidirectional attention mask, which enables the model to capture essential user traits across global sequence representations. 
In the same way, the outputs of the last layer are also divided into two scales:
\begin{align}
    \bm{f}_{u_i} = \bm{H}^{L}_{0}, \ \ \ \ \
    \widetilde{\bm{f}}_{u_i} = \bm{H}^{L}_{1:n+1},
\end{align}
where $\bm{f}_{u_i} \in \mathbb{R}^{d}$ is the user's macro passive dynamic representation, and $\widetilde{\bm{f}}_{u_i} \in \mathbb{R}^{n \times d}$ is the user's micro passive dynamic representations.

For user $v_{j}$ on the other side, we also encode his active and passive dynamic representations based on the symmetrical model structure, which are denoted by $\bm{p}_{v_j}$, $\widetilde{\bm{p}}_{v_j}$ and $\bm{f}_{v_j}$, $\widetilde{\bm{f}}_{v_j}$.

\subsection{Multi-Scale Sequence Matching} \label{sec:match}
The above process allows us to obtain multi-scale active and passive dynamic representations of users by encoding their behavior sequences. In the following section, we will delve into the dual-perspective multi-scale matching of both users in reciprocal recommendation.
\subsubsection{Macro-Level Matching}
As mentioned in the previous sections, the matching of both parties in reciprocal recommendation is a dual-perspective active selection process~\cite{pizzato2010recon,xia2015reciprocal,neve2019latent,yang2022modeling}. Hence, we apply two scores for active-passive matching to model this dual-perspective propensity. It is worth noting that in Figure~\ref{fig:model}, only one perspective matching process is depicted, while the other perspective is symmetrical. The matching prediction method at the macro level is as follows:
\begin{align}
\begin{split}
    {\hat{y}}_{i,j} & = {\hat{y}}_{i \rightarrow j} + {\hat{y}}_{j \rightarrow i}\\
    &= (\bm{p}_{u_i})^{T} \cdot \bm{f}_{v_j} + (\bm{p}_{v_j})^{T} \cdot \bm{f}_{u_i},
\end{split}
\label{eq:ma_score}
\end{align}
where $\cdot$ is the dot product operation. The first part of the formula represents the propensity from $u_i$ to $v_j$, and the second part represents the propensity from $v_j$ to $u_i$.
\subsubsection{Time-Sensitive Micro-Level Matching}
Compared with the conventional recommender systems that rely on a fixed number of static items as candidate-matching targets, our problem formulation involves a distinctive sequence matching task, which employs dynamic user behavior sequences as candidate-matching targets. This results in a broader and more dispersed candidate matching space. 
Consequently, to alleviate the sparsity of the candidate matching space, we strive to attain more discriminative and precise matching predictions by incorporating more detailed and fine-grained micro-level interactions between users on both sides. 
Formally, the overall micro-level matching method is as follows:
\begin{align}
    {\hat{z}}_{i,j} = \operatorname{TiSensiMatch} \left(\widetilde{\bm{p}}_{u_i}, \ \widetilde{\bm{f}}_{v_j}\right) + \operatorname{TiSensiMatch} \left(\widetilde{\bm{p}}_{v_j}, \  \widetilde{\bm{f}}_{u_i}\right).
    \label{eq:mi_score}
\end{align}

Next, as shown in the ``Time-Sensitive Micro-Level Matching'' section of Figure~\ref{fig:model}, we take $\operatorname{TiSensiMatch} \left(\widetilde{\bm{p}}_{u_i}, \ \widetilde{\bm{f}}_{v_j}\right)$ as an example to describe the implementation of time-sensitive matching in detail. 
Firstly, we compute the fine-grained matching matrix between the micro active dynamic representations $\widetilde{\bm{p}}_{u_i}$ and the passive dynamic representations $\widetilde{\bm{f}}_{v_j}$:
\begin{align}
    \bm{G} = \widetilde{\bm{p}}_{u_i} \cdot {(\widetilde{\bm{f}}_{v_j})}^{T},
\end{align}
where $\bm{G} \in \mathbb{R}^{n \times n}$ stands for the matching degrees between the active and passive representations at different time steps. 
Secondly, we aggregate the matrix $\bm{G}$ through two dimensions of co-attention. For the passive dimension, we compute attention weights based on softmax:
\begin{align}
    \bm{\gamma} = \operatorname{softmax}(\widetilde{\bm{f}}_{v_j} \cdot \bm{e}^{p}_{u_i}),
\end{align}
where $\bm{e}^{p}_{u_i} \in \mathbb{R}^{d}$ is the original active embedding of the user $u_i$ from $\bm{M}^{U}_{p}$. For the active dimension, we propose a relative time-dependent attention weight to model the time-sensitivity of user active dynamic behaviors:
\begin{align}
    \bm{\delta} = \operatorname{softmax}(\widetilde{\bm{p}}_{u_i} \cdot \bm{e}^{f}_{v_j} + \bm{\alpha}^{\tau}),
\end{align}
where $\bm{e}^{f}_{v_j} \in \mathbb{R}^{d}$ is the original passive embedding of the user $v_j$ from $\bm{M}^{V}_{f}$. 
And $\bm{\alpha}^{\tau}$ is from a learnable relative time weight $\bm{\alpha} \in \mathbb{R}^{n}$ starting from the end of the sequence. 
For a behavior sequence with an actual length of $s$ (no padding), there is $\bm{\alpha}^{\tau}_{s-k} = \bm{\alpha}_{k-1}$, where $k\in\{1,2,\ldots,s\}$. 
Finally, we aggregate the fine-grained matching matrix $\bm{G}$ in the following way:
\begin{align}
\begin{split}
     {\hat{z}}_{i \rightarrow j} &= \operatorname{TiSensiMatch} \left(\widetilde{\bm{p}}_{u_i}, \ \widetilde{\bm{f}}_{v_j}\right) \\
    & =  {\bm{\delta}}^{T} \cdot \bm{G} \cdot \bm{\gamma}.
\end{split}
\label{eq:agg}
\end{align}

In addition, for the perspective from $v_j$ to $u_i$, we also get the micro propensity score ${\hat{z}}_{j \rightarrow i}$ in the same way. The full micro-level matching score ${\hat{z}}_{i,j}$ is the sum of the dual-perspective propensities.

\subsection{Improving Matching Efficiency via Self-Distillation}
The rapid development of the internet results in an increased number of users on online platforms. However, this also leads to a service efficiency problem that needs to be solved urgently. 
Specific to the multi-scale sequence matching method we proposed, macro-level matching is based on two simple and efficient dot product operations, while micro-level matching provides more fine-grained bilateral interactions. 
Nevertheless, the complexity of the latter also presents efficiency issues that need to be addressed.

Inspired by knowledge distillation techniques~\cite{hinton2015distilling} that have been successfully applied in various fields~\cite{zhang2019your,sanh2019distilbert,hofstatter2020improving,tian2023directed}, we also aspire to improve the recommendation efficiency in practical applications through self-distillation from micro-level matching to macro-level matching.
More exactly, as shown in the ``Micro-to-Macro Self-Distillation'' section of Figure~\ref{fig:model}, we use the self-distillation technique to transfer knowledge from finer-grained micro-level matching to more efficient and simple macro-level matching during training. 
Furthermore, only macro-level matching is used in the process of verification, testing and practical application to achieve efficient and precise recommendations.

\paratitle{Micro-to-macro self-distillation.}
As a ranking task, reciprocal recommendation differs from common classification problems in that it pays more attention to the relative score gap between positive and negative instances rather than the absolute category distribution. Therefore, we adopt Margin-MSE~\cite{hofstatter2020improving} as our distillation loss to facilitate the alignment of positive and negative score margins between micro-level and macro-level matching. Specifically, the calculation for self-distillation loss is:
\begin{align}
\begin{split}
\mathcal{L}_{sd}
    & = \operatorname{Margin-MSE} \left({\hat{z}}^{+}, \ {\hat{z}}^{-}, \ {\hat{y}}^{+}, \ {\hat{y}}^{-}\right), \\
    & = \frac{1}{|B|} \sum^{|B|}_{i=1} \left[(\hat{z}^{+}_{i} - \hat{z}^{-}_{i}) - (\hat{y}^{+}_i -\hat{y}^{-}_i )\right]^{2},
\end{split}
\end{align}
where $|B|$ represents the batch size of training data. Besides, ${\hat{z}}^{+}$, ${\hat{z}}^{-}$ and ${\hat{y}}^{+}$, ${\hat{y}}^{-}$ are the positive and negative score sets of micro level and macro level, respectively. 

\paratitle{Overall optimization.}
In our approach, the final matching prediction is computed by four elements from two users (Eqn.~\eqref{eq:ma_score} and ~\eqref{eq:mi_score}), which are indispensable in the actual scenario for successful matching. Thus, we can get four negative instance scores by replacing the four elements with negative instances, respectively. Formally, for each positive instance $(u_i, {S}^T_{u_i}, v_j, {S}^T_{v_j})$, we separately sample bilateral negative users $(u_{i'}, {S}^T_{u_{i'}})$ and $(v_{j'}, {S}^T_{v_{j'}})$. Note that the historical behavior sequences of negative users are also truncated to the current time $T$. After inputting negative instances into the model, four sets of negative instances at the macro level and micro level can be obtained:
\begin{align}
    \bm{D} = \{(\bm{p}_{u_{i'}}, \widetilde{\bm{p}}_{u_{i'}}),(\bm{f}_{u_{i'}}, \widetilde{\bm{f}}_{u_{i'}}),(\bm{p}_{v_{j'}}, \widetilde{\bm{p}}_{v_{j'}}),(\bm{f}_{v_{j'}}, \widetilde{\bm{f}}_{v_{j'}})\}.
\end{align}

\textcolor{black}{By replacing the corresponding elements in Eqn.~\eqref{eq:ma_score} and ~\eqref{eq:mi_score} with the aforementioned negative instances, we can compute multiple negative instance scores for both scales. Without loss of generality, we denote them as two negative instance score sets, $\bm{Y}^{-}_{i,j}$ and $\bm{Z}^{-}_{i,j}$, each containing four types of negative instance scores.} 
Then, we use the BPR loss~\cite{rendle2009bpr} to optimize the ranking between positive and negative instances:
\begin{align}
    \mathcal{L}_{ma}
    & = - \frac{1}{|B|} \sum_{(u_i,v_j) \in B} \ \sum_{\hat{y}^{-}_{i,j} \in \bm{Y}^{-}_{i,j}} \log \left( \sigma ( \hat{y}^{+}_{i,j}-\hat{y}^{-}_{i,j})\right), \\
    \mathcal{L}_{mi}
    & = - \frac{1}{|B|} \sum_{(u_i,v_j) \in B} \ \sum_{\hat{z}^{-}_{i,j} \in \bm{Z}^{-}_{i,j}} \log \left( \sigma ( \hat{z}^{+}_{i,j}- \hat{z}^{-}_{i,j} )\right),
\end{align}
where $B$ represents a batch of training data. $\hat{y}^{+}_{i,j}$ refers to the macro-level matching score of user pairs, and $\hat{y}^{-}_{i,j}$ denotes its negative instance score. $\hat{z}^{+}_{i,j}$ and $\hat{z}^{-}_{i,j}$ are corresponding to micro-level matching. In addition, based on multiple negative instance scores, the self-distillation loss can also be extended to:
\begin{align}
    \mathcal{L}_{sd}
     = \frac{1}{|B|} \sum_{(u_i,v_j) \in B}  \sum_{\scriptscriptstyle{\hat{y}^{-}_{i,j} \in \bm{Y}^{-}_{i,j}, \hat{z}^{-}_{i,j} \in \bm{Z}^{-}_{i,j}}} \scriptstyle{\left[(\hat{z}^{+}_{i,j} - \hat{z}^{-}_{i,j}) - (\hat{y}^{+}_{i,j} -\hat{y}^{-}_{i,j})\right]^{2}}.
     \label{eq:lsd}
\end{align}

Combining the above three losses, our overall optimization loss function can be denoted as follows: 
\begin{align}
    \mathcal{L} = \mathcal{L}_{ma} + \lambda \mathcal{L}_{mi} + \mu \mathcal{L}_{sd},
\end{align}
where $\lambda$ and $\mu$ are hyperparameters. $\lambda$ is used to balance the training of macro and micro, while $\mu$ is for the trade-off between ranking optimization and self-distillation knowledge transfer.

Note that during validation and testing, we only deploy macro-level matching module, which makes the model much more efficient in practice.

\subsection{Discussion}
\subsubsection{Time Complexity Analysis}
In a real-world deployment, the model inference associated only with the entities themselves can be completed in advance.
In reference to our approach, the encoding of a user's behavior sequence is independent of other individuals, and the outcomes can be pre-calculated and stored. 
Therefore, we mainly discuss the time complexity of the matching prediction part.
Matching at the macro level requires only the time complexity of $\mathcal{O}(d)$, where $d$ denotes the model output dimension. It is the same as many classical collaborative filtering-based methods~\cite{rendle2009bpr,neve2019latent} and sequential recommendation methods~\cite{kang2018self,sun2019bert4rec}.
For micro-level matching, it first takes $\mathcal{O}(n^2d)$ time to complete the calculation of the fine-grained matching matrix, where $n$ denotes the sequence length. Then, the time complexity for the co-attention of the two dimensions is $\mathcal{O}(nd)$. The final aggregation operation still takes time of $\mathcal{O}(n^2)$. Overall, the time complexity of prediction at the micro level is $\mathcal{O}(n^2d)$.
Given that online platforms cater to tens of thousands of users, such time complexity is unacceptable. As a result, we suggest utilizing self-distillation from the micro level to the macro level to optimize the model's efficiency for practical application.
\subsubsection{Comparison with Existing Recommendation Methods}
Here, we briefly compare with related recommendation methods to highlight the innovation and novelty of our approach.

\paratitle{Sequential recommendation methods.}
Existing sequential recommendation algorithms such as SASRec~\cite{kang2018self} and BERT4Rec~\cite{sun2019bert4rec} only encode user dynamic representations from one perspective, which lacks the overall modeling of dual perspectives in the reciprocal recommendation scenario. In contrast, the methodology we proposed involves the dual-perspective joint modeling of users. And we achieve more precise recommendations by performing fine-grained matching on more details of both recommending parties.

\paratitle{Reciprocal recommendation methods.}
In the current reciprocal recommendation studies (regardless of content-based or collaborative filtering methods), the modeling of the user ultimately results in a static representation vector. 
Some methods~\cite{pizzato2010recon,yildirim2021bideepfm} model user preferences based on their fixed profiles, while others~\cite{rendle2009bpr,neve2019latent} learn user's static implicit representations through matrix factorization and latent factor models. 
Even methods that are also based on user historical behaviors~\cite{fu2021beyond} only aggregate all user histories in the training set, while ignoring the temporal dynamics of behavior sequences. 
As a comparison, our method takes into full consideration the dynamics on both user active and passive perspectives when constructing model inputs (Section~\ref{sec:pb_fmu}), encoding user behaviors (Section~\ref{sec:bhv_model}), and predicting bilateral user matching (Section~\ref{sec:match}).

\section{Experiment}

\subsection{Experiment Setup}
\subsubsection{Dataset}
We evaluate our model in two reciprocal recommendation scenarios encompassing five real-world datasets.
\begin{enumerate}
\item \textbf{Online recruitment.} We conduct the evaluation using three authentic datasets sourced from a popular Chinese online recruitment platform. These datasets are created from over 100 days of real online logs, with all user private information removed. 
The employer-employee pairs that achieved an interview are gathered as instances of positive interactions.
\item \textbf{Question answering.}
We use two public datasets from distinct sections of the popular online Q\&A website ``Stack Exchange''~\cite{paranjape2017motifs}, intended for mutual matching between questioners and answerers.
For the \textit{answerer} side, we recommend questions to users according to their historical answering records. For the \textit{questioner} side, we follow an expert finding~\cite{lin2017survey} paradigm to recommend expert users for questioners.
\end{enumerate}

Regarding data processing, we initially remove inactive users through 5-core filtering and then create a behavior sequence for each user based on chronological order. For each interaction record, we only use the sequence truncated to the current interaction time to prevent data leakage during training.
On the three recruitment datasets, we adopt the last two weeks of data as the \emph{validation} and \emph{test} sets, and the remaining data is used for \emph{training}. Besides, the datasets for questioner-answerer matching are divided based on the time span in an 8:1:1 ratio.
The overall statistics are shown in Table~\ref{tab:data_statistics}.
Our code is available at \textcolor{blue}{\url{https://github.com/RUCAIBox/ReSeq/}}.

\begin{table}[]
\caption{Statistics of the experimental datasets.}
\label{tab:data_statistics}
\resizebox{0.45\textwidth}{!}{
\setlength{\tabcolsep}{1.0mm}{
\begin{tabular}{@{}ccrrrr@{}}
\toprule
\multirow{4}{*}{\textbf{Recruitment}}
 & \textbf{Datasets} & \textbf{\#Candidates} & \textbf{\#Recruiters}  & \textbf{\#Interactions} & \textbf{Sparsity} \\ \cmidrule(l){2-6}
& Design & 12,274 & 9,141  & 139,355        & 99.88\%  \\
                             & Sale & 15,831 & 12,757 & 112,340        & 99.94\%  \\
                             & Technology & 56,620 & 48,071 & 808,376 & 99.97\%  \\ \midrule
\multirow{3}{*}{\textbf{Q\&A}}        
& \textbf{Datasets} & \textbf{\#Questioners} & \textbf{\#Answerers} & \textbf{\#Interactions} & \textbf{Sparsity} \\ \cmidrule(l){2-6}
& StackOverflow & 42,381 & 25,004  & 360,063                                     & 99.97\%  \\
                             & AskUbuntu & 6,030 & 3,415 & 69,865        & 99.66\% \\ \bottomrule
\end{tabular}}}
\end{table}

\begin{table*}[]
\caption{Performance comparison of different methods on the three datasets in the recruitment scenario. The best-performed and the second-best-performed methods are indicated in bold and underlined font, respectively. ``*'' denotes that the improvements are significant at the level of 0.01 with paired $t$-test.}
\label{tab:recruit_comparison}
\resizebox{0.82\linewidth}{!}{
\setlength{\tabcolsep}{2.7mm}{
\begin{tabular}{@{}cccccccc@{}}
\toprule
\multirow{2}{*}{Dataset}     & Perspective & \multicolumn{3}{c}{Candidates}                         & \multicolumn{3}{c}{Recruiters}                      \\ \cmidrule(l){2-2} \cmidrule(l){3-5} \cmidrule(l){6-8}
                             & Metric      & HR@5            & MRR@5           & NDCG@5          & HR@5            & MRR@5           & NDCG@5          \\ \midrule
\multirow{13}{*}{Design}     & BPR         & 0.1437          & 0.0866          & 0.1008          & 0.1300            & 0.0693          & 0.0842          \\
                             & LFRR        & 0.1465          & 0.0855          & 0.1006          & 0.1338          & 0.0694          & 0.0853          \\
                             & NeuMF       & 0.1611          & 0.1016          & 0.1163          & 0.1452          & 0.0804          & 0.0964          \\
                             & LightGCN    & 0.1496          & 0.0913          & 0.1057          & 0.1349          & 0.0732          & 0.0884          \\ \cmidrule(l){2-8}
                             & SASRec      & 0.2033          & 0.1184          & 0.1394          & 0.2529          & 0.1389          & 0.1670           \\
                             & SSE-PT      & 0.1961 & 0.1097	& 0.1311	& 0.2466	& 0.1327	& 0.1608                     \\
                             & BERT4Rec    & 0.1535          & 0.0874          & 0.1037          & 0.2027          & 0.1044          & 0.1286          \\
                             & FMLP-Rec     & 0.1940           & 0.1194          & 0.1379          & \underline{0.2845}    & \underline{0.1571}    & \underline{0.1886}    \\ \cmidrule(l){2-8}
                             & PJFNN       & 0.1056          & 0.0621          & 0.0728          & 0.0604          & 0.0272          & 0.0353          \\
                             & BPJFNN      & 0.1755          & 0.1010           & 0.1194          & 0.1589          & 0.0766          & 0.0968          \\
                             & IPJF        & 0.1778          & 0.0943          & 0.1149          & 0.1661          & 0.0835          & 0.1038          \\
                             & PJFFF       & 0.1839          & 0.1052          & 0.1247          & 0.1235          & 0.0595          & 0.0752          \\
                             & DPGNN       & \underline{0.2422}    & \underline{0.1396}    & \underline{0.1649}    & 0.2317          & 0.1273          & 0.1530           \\ \cmidrule(l){2-8}
                             & ReSeq        & \textbf{0.4435*} & \textbf{0.2554*} & \textbf{0.3020*}  & \textbf{0.3722*} & \textbf{0.1994*} & \textbf{0.2420*}  \\ \midrule
\multirow{13}{*}{Sale}       & BPR         & 0.1071          & 0.0613          & 0.0726          & 0.0929          & 0.0513          & 0.0615          \\
                             & LFRR        & 0.1262          & 0.0730           & 0.0862          & 0.1083          & 0.0602          & 0.0721          \\
                             & NeuMF       & 0.1367          & 0.0828          & 0.0962          & 0.1113          & 0.0639          & 0.0756          \\
                             & LightGCN    & 0.1447          & 0.0832          & 0.0984          & 0.1228          & 0.0691          & 0.0823          \\ \cmidrule(l){2-8}
                             & SASRec      & 0.1707          & 0.0932          & 0.1123          & 0.1494          & 0.0807          & 0.0976          \\
                             & SSE-PT      & 0.1592	& 0.0912	& 0.1080	& 0.1276	& 0.0648	& 0.0802 \\
                             & BERT4Rec    & 0.1303          & 0.0685          & 0.0837          & 0.1143          & 0.0579          & 0.0718          \\
                             & FMLP-Rec     & 0.1758          & 0.1001          & 0.1188          & 0.1384          & 0.0742          & 0.0901          \\ \cmidrule(l){2-8}
                             & PJFNN       & 0.1097          & 0.0650           & 0.0760           & 0.0568          & 0.0269          & 0.0342          \\
                             & BPJFNN      & 0.1355          & 0.0767          & 0.0912          & 0.1130           & 0.0549          & 0.0692          \\
                             & IPJF        & 0.1092          & 0.0541          & 0.0677          & 0.1177          & 0.0629          & 0.0763          \\
                             & PJFFF       & 0.1564          & 0.0876          & 0.1046          & 0.1052          & 0.0524          & 0.0654          \\
                             & DPGNN       & \underline{0.1871}    & \underline{0.1026}    & \underline{0.1234}    & \underline{0.1703}    & \underline{0.0930}     & \underline{0.1121}    \\ \cmidrule(l){2-8}
                             & ReSeq        & \textbf{0.2712*} & \textbf{0.1499*} & \textbf{0.1798*} & \textbf{0.2434*} & \textbf{0.1335*} & \textbf{0.1606*} \\ \midrule
\multirow{13}{*}{Technology} & BPR         & 0.2733          & 0.1778          & 0.2015          & 0.2659          & 0.1627          & 0.1883          \\
                             & LFRR        & 0.2270           & 0.1420           & 0.1631          & 0.2151          & 0.1232          & 0.1459          \\
                             & NeuMF       & 0.2796          & 0.1821          & 0.2063          & 0.2711          & 0.1662          & 0.1922          \\
                             & LightGCN    & 0.2775          & 0.1823          & 0.2060           & 0.2726          & 0.1695          & 0.1952          \\ \cmidrule(l){2-8}
                             & SASRec      & 0.4145          & 0.2815          & 0.3148          & 0.4702          & 0.3042          & 0.3456          \\
                             & SSE-PT      & 0.3924	& 0.2595	& 0.2926   & 0.4931	& 0.3129	& 0.3578\\
                             & BERT4Rec    & 0.3650           & 0.2408          & 0.2718          & 0.4243          & 0.2718          & 0.3098          \\
                             & FMLP-Rec     & 0.4165          & \underline{0.2898}    & \underline{0.3215}    & \underline{0.5206}    & \underline{0.3476}    & \underline{0.3908}    \\ \cmidrule(l){2-8}
                             & PJFNN       & 0.2707          & 0.1478          & 0.1781          & 0.2636          & 0.1348          & 0.1666          \\
                             & BPJFNN      & 0.3791          & 0.2172          & 0.2573          & 0.4261          & 0.2273          & 0.2764          \\
                             & IPJF        & 0.3403          & 0.1743          & 0.2151          & 0.3776          & 0.1996          & 0.2435          \\
                             & PJFFF       & 0.3425          & 0.2126          & 0.2448          & 0.3715          & 0.1998          & 0.2422          \\
                             & DPGNN       & \underline{0.4521}    & 0.2696          & 0.3148          & 0.4409          & 0.2599          & 0.3047          \\ \cmidrule(l){2-8}
                             & ReSeq        & \textbf{0.7597*} & \textbf{0.4945*} & \textbf{0.5606*} & \textbf{0.7809*} & \textbf{0.5099*} & \textbf{0.5775*} \\ \bottomrule
\end{tabular}}}
\end{table*}

\begin{table*}[]
\caption{Performance comparison of different methods on two datasets in the questioner-answerer matching scenario. The best-performed and the second-best-performed methods are indicated in bold and underlined font, respectively. ``*'' denotes that the improvements are significant at the level of 0.01 with paired $t$-test.}
\label{tab:qa_comparison}
\resizebox{0.82\linewidth}{!}{
\setlength{\tabcolsep}{2.7mm}{
\begin{tabular}{@{}cccccccc@{}}
\toprule
\multirow{2}{*}{Dataset}       & Perspective & \multicolumn{3}{c}{Questioners}                     & \multicolumn{3}{c}{Answerers}                       \\ \cmidrule(l){2-2} \cmidrule(l){3-5} \cmidrule(l){6-8}
 & Metric   & HR@5   & MRR@5  & NDCG@5 & HR@5   & MRR@5  & NDCG@5 \\ \midrule
\multirow{8}{*}{StackOverflow} & BPR         & 0.3262          & 0.2344          & 0.2572          & 0.2653          & 0.1825          & 0.2029          \\
 & LFRR     & 0.3051 & 0.2113 & 0.2346 & 0.2391 & 0.1522 & 0.1737 \\
 & NeuMF    & 0.3218 & 0.2351 & 0.2567 & 0.2563 & 0.1781 & 0.1975 \\
 & LightGCN & 0.3352 & 0.2359 & 0.2606 & 0.2780  & 0.1881 & 0.2103 \\ \cmidrule(l){2-8}
 & SASRec   & 0.4336 & 0.3193 & 0.3478 & 0.2672 & 0.1780  & 0.2001 \\
 & SSE-PT   & 0.4064	& 0.2901	& 0.3190	& 0.2857	& 0.1881	& 0.2123  \\
 & BERT4Rec & 0.3621 & 0.2501 & 0.2779 & 0.2456 & 0.1548 & 0.1772 \\
                               & FMLP-Rec     & \underline{0.4594}    & \textbf{0.3428} & \textbf{0.3718} & \underline{0.2957}    & \underline{0.2026}    & \underline{0.2257}    \\ \cmidrule(l){2-8}
                               & ReSeq        & \textbf{0.4868*} & \underline{0.3235}    & \underline{0.3640}     & \textbf{0.4019*} & \textbf{0.2432*} & \textbf{0.2825*} \\ \midrule
\multirow{8}{*}{AskUbuntu}     & BPR         & 0.1566          & 0.1106          & 0.1220           & 0.1299          & 0.0914          & 0.1009          \\
 & LFRR     & 0.1590  & 0.1003 & 0.1149 & 0.0935 & 0.0622 & 0.0699 \\
 & NeuMF    & 0.2351 & 0.1592 & 0.1780  & 0.1323 & 0.0992 & 0.1074 \\
 & LightGCN & 0.1606 & 0.1128 & 0.1247 & 0.1245 & 0.0906 & 0.0990 \\ \cmidrule(l){2-8}
  & SASRec   & 0.2663 & 0.1698 & 0.1937 & 0.1588 & 0.1076 & 0.1202 \\
  & SSE-PT   & 0.2418	& 0.1584	& 0.1791	& 0.1540	& 0.0994	& 0.1129 \\
 & BERT4Rec & 0.2294 & 0.1404 & 0.1626 & 0.1143 & 0.0669 & 0.0786 \\
                               & FMLP-Rec     & \underline{0.2706}    & \underline{0.1818}    & \underline{0.2039}    & \underline{0.1627}    & \underline{0.1121}    & \underline{0.1246}    \\ \cmidrule(l){2-8}
                               & ReSeq        & \textbf{0.5259*} & \textbf{0.2848*} & \textbf{0.3447*} & \textbf{0.2254*} & \textbf{0.1455*} & \textbf{0.1653*} \\
                               \bottomrule
\end{tabular}}}
\end{table*}

\subsubsection{Baseline Models}
We compare ReSeq with the following baseline models (key configurations are in parentheses):
\begin{itemize}
\item {\textbf{BPR}~\cite{rendle2009bpr}} is a matrix factorization-based collaborative filtering method, which is optimized by Bayesian Personalized Ranking (BPR) loss (embedding size=64).
\item {\textbf{LFRR}~\cite{neve2019latent}} applies the latent factor model for reciprocal recommendation, which utilizes two latent factor models to model the dual perspectives of both parties (embedding size=64).
\item {\textbf{NeuMF}~\cite{he2017neural}} combines matrix factorization and neural networks to improve the effectiveness of personalized recommendation (MLP hidden sizes= [128,64], dropout ratio=0.1).
\item {\textbf{LightGCN}~\cite{he2020lightgcn}} is a collaborative filtering method based on lightweight graph convolutional networks, which simplifies graph convolution operations and achieves competitive performance on user-item recommendation (embedding size=64, number of layers=2).
\item {\textbf{SASRec}~\cite{kang2018self}} exploits a unidirectional Transformer-based neural network to model the item sequences and predict the next item (number of layers=2, number of attention heads=2, dropout ratio=0.5).
\item {\textbf{SSE-PT}~\cite{wu2020sse}} combines user and item embeddings as input of the Transformer model for personalized sequential recommendation (number of layers=2, number of attention heads=2, dropout ratio=0.5).
\item {\textbf{BERT4Rec}~\cite{sun2019bert4rec}} adopts a bidirectional Transformer model and combines it with a mask prediction task for the modeling of item sequences (number of layers=2, number of attention heads=2, fine-tuning ratio=0.5)
\item {\textbf{FMLP-Rec}~\cite{zhou2022filter}} proposes an all-MLP model with learnable filters, which ensures efficiency and reduces noisy signals (number of layers=2, dropout ratio=0.5).
\item {\textbf{PJFNN}~\cite{zhu2018person}} is a person-job fit model based on CNN to predict the matching degree between resumes and job descriptions (max sentence number=20, max sentence length=30).
\item {\textbf{BPJFNN}~\cite{qin2018enhancing}} leverages a bidirectional LSTM network to encode resumes and job descriptions for person-job fit prediction (max sentence length=512, number of layers=1, hidden size=32, dropout ratio=0.1)
\item {\textbf{IPJF}~\cite{le2019towards}} learns the matching propensities between candidates and recruiters based on various behavior labels in recruitment scenarios (max sentence number=20, max sentence length=30, BERT embedding size=768).
\item {\textbf{PJFFF}~\cite{jiang2020learning}} fuses the user explicit and implicit representations. The implicit representations are modeled by LSTM based on historical user records (hidden size=32, historical item length=100, BERT embedding size=768, BERT output size=32).
\item {\textbf{DPGNN}~\cite{yang2022modeling}} proposes a dual-perspective graph representation learning method that models recruitment as a two-way selection process (number of layers=2, BERT embedding size=768, BERT output size=32, contrastive loss weight=0.05, temperature=0.2).
\end{itemize}
Our baselines can be categorized into collaborative filtering models (\ie BPR, LFRR, NeuMF, LightGCN), sequential recommendation models (\ie SASRec, SSE-PT, BERT4Rec, FMLP-Rec), and person-job fit models (\ie PJFNN, BPJFNN, IPJF, PJFFF, DPGNN). Among these, the person-job fit models are exclusively applicable in online recruitment scenarios where users possess comprehensive textual information.

\subsubsection{Evaluation and Implementation Details}
Following ~\cite{yang2022modeling}, we employ three widely used metrics, Hit Ratio (HR@$k$), Normalized Discounted Cumulative Gain (NDCG@$k$), and Mean Reciprocal Rank (MRR@$k$) to evaluate ranking tasks from dual perspectives simultaneously, where $k$ is set to 5.
Concretely, for each positive instance interaction $(u_i, {S}^T_{u_i}, v_j, {S}^T_{v_j})$, we randomly sample $100$ users from $\mathcal{U}$ and $\mathcal{V}$ as dual-perspective negative instances, respectively. Then truncate their behavior sequences to the current interaction time for the evaluation to prevent data leakage~\cite{zhao2022revisiting}.

For all the compared methods, we use Adam for optimization and grid search for hyperparameter tuning to get optimal performance. The learning rates of all models are tuned in \{0.005, 0.001, 0.0005, 0.0003, 0.0001, 0.00001\}.
To ensure a fair comparison, the dimensions of embeddings are uniformly set to $64$. Besides, the weight decay coefficient is set to $1e-5$. And we select the best models by early stopping with the patience of $10$ epochs.
All models are implemented based on a popular open-source recommendation library \textsc{RecBole}~\cite{zhao2021recbole,zhao2022recbole}.

\begin{table*}[]
\caption{Ablation study of our method in two scenarios. The best-performed methods are indicated in bold.}
\label{tab:ablation_study}
\resizebox{0.85\linewidth}{!}{
\setlength{\tabcolsep}{2.5mm}{
\begin{tabular}{@{}clcccccc@{}}
\toprule
\multirow{2}{*}{Dataset} &
  Perspective &
  \multicolumn{3}{c}{Candidates/Questioners} &
  \multicolumn{3}{c}{Recruiters/Answerers} \\ \cmidrule(l){2-2} \cmidrule(l){3-5} \cmidrule(l){6-8}
 & 
  Metric &
  \multicolumn{1}{c}{HR@5} &
  \multicolumn{1}{c}{MRR@5} &
  \multicolumn{1}{c}{NDCG@5} &
  \multicolumn{1}{c}{HR@5} &
  \multicolumn{1}{c}{MRR@5} &
  \multicolumn{1}{c}{NDCG@5} \\ \midrule
\multirow{5}{*}{Design} &
  ReSeq &
  \textbf{0.4435} &
  \textbf{0.2554} &
  \textbf{0.3020} &
  \textbf{0.3722} &
  \textbf{0.1994} &
  \textbf{0.2420} \\
 & \ \ w/o DSE & 0.4224 & 0.2462 & 0.2898 & 0.3409 & 0.1806 & 0.2201 \\
 & \ \ w/o MASK & 0.4305 & 0.2510  & 0.2954 & 0.3406 & 0.1836 & 0.2223 \\
 & \ \ w/o TSA & 0.4139 & 0.2368 & 0.2806 & 0.3315 & 0.1798 & 0.2172 \\
 & \ \ w/o SD   & 0.4156 & 0.2372 & 0.2813 & 0.3446 & 0.1889 & 0.2273 \\ \midrule
\multirow{5}{*}{StackOverflow} &
  ReSeq &
  \textbf{0.4868} &
  \textbf{0.3235} &
  \textbf{0.3640} &
  \textbf{0.4019} &
  \textbf{0.2432} &
  \textbf{0.2825} \\
 & \ \ w/o DSE  & 0.4787 & 0.3102 & 0.3521 & 0.3964 & 0.2391 & 0.2780  \\
 & \ \ w/o MASK & 0.4764 & 0.3114 & 0.3524 & 0.3942 & 0.2387 & 0.2772 \\
 & \ \ w/o TSA & 0.4710  & 0.3128 & 0.3521 & 0.3697 & 0.2239 & 0.2600   \\
 & \ \ w/o SD   & 0.4787 & 0.3154 & 0.3560  & 0.3935 & 0.2377 & 0.2763 \\ \bottomrule
\end{tabular}}}
\end{table*}

\begin{figure*}[]
	\centering
	\includegraphics[width=0.495\linewidth]{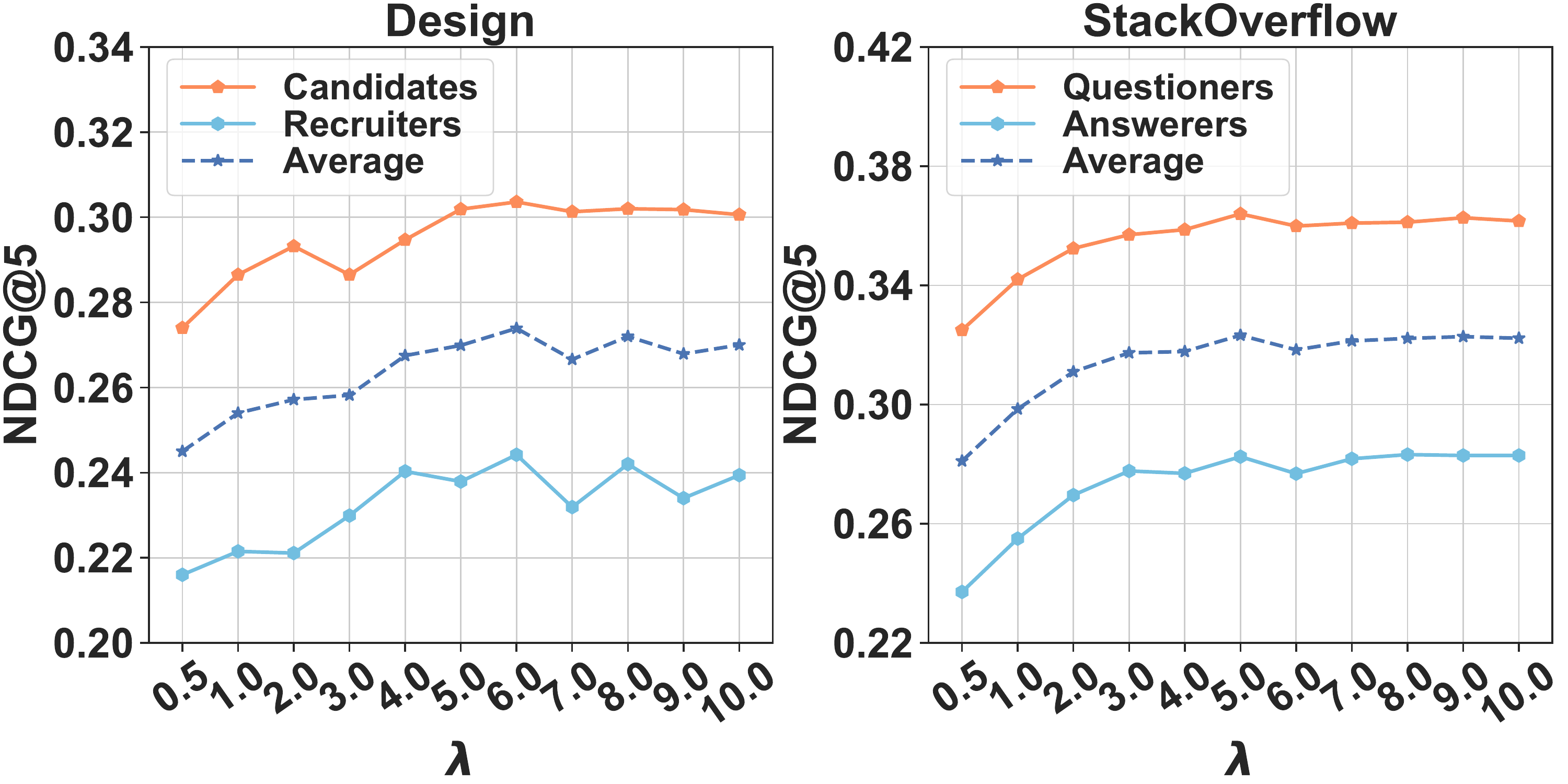}
 \includegraphics[width=0.495\linewidth]{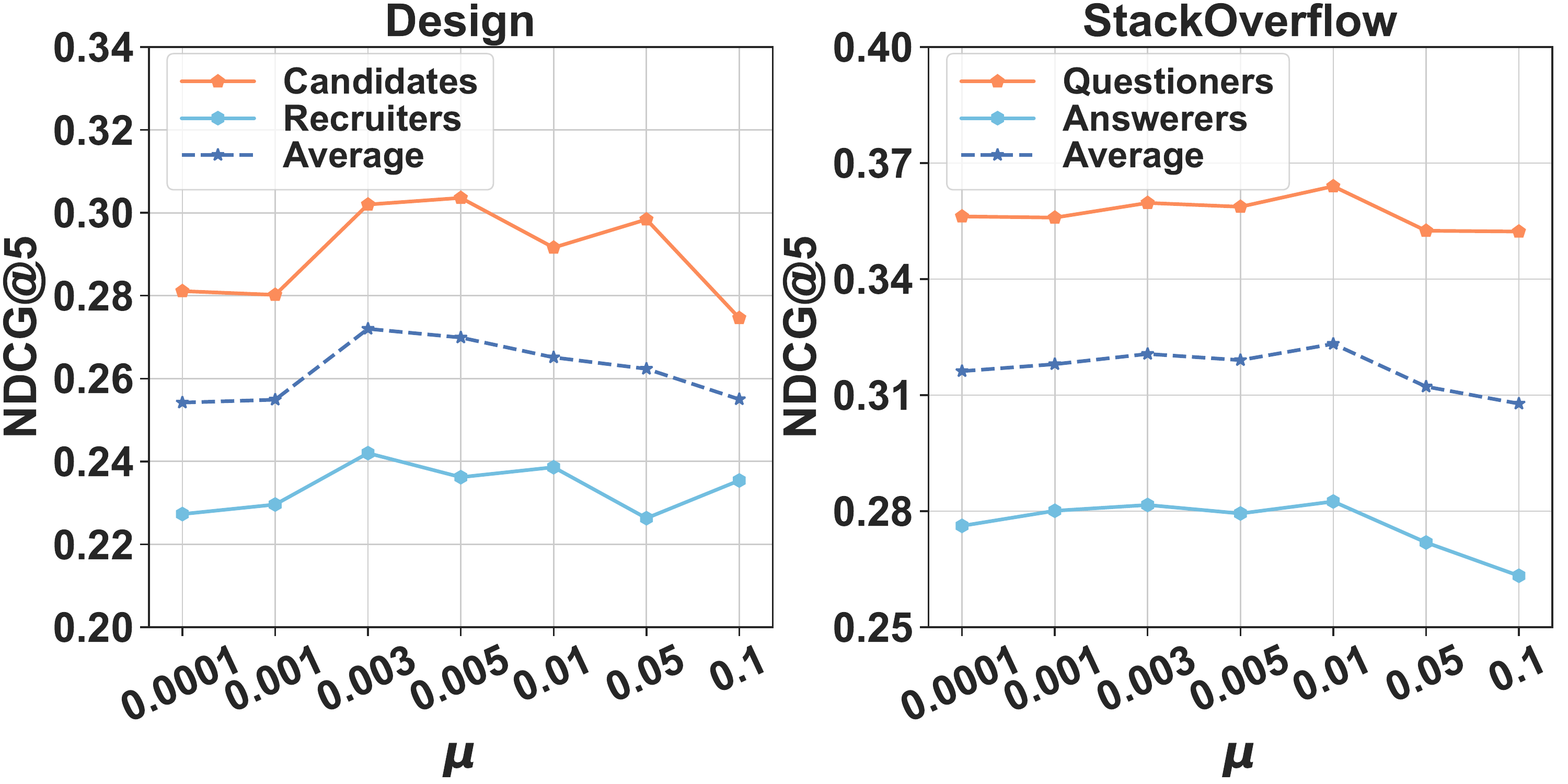}
	\caption{Performance comparison w.r.t. micro teacher loss coefficient $\lambda$ and self-distillation loss coefficient $\mu$.}
	\label{fig:parameters}
\end{figure*}

\subsection{Experiment Results}
The results of different methods in the recruitment scenario are shown in Table~\ref{tab:recruit_comparison}, including baseline models dedicated to person-job fit (with extra text data). Table~\ref{tab:qa_comparison} shows the comparison between our model and baseline models in the questioner-answerer matching scenario. Based on these results, we can find:

For the four collaborative filtering methods, models based on GCN (\ie LightGCN) or MLP (\ie NeuMF) for hybrid interaction tend to exhibit superior performance in comparison to simple matrix factorization method (\ie BPR) and latent factor model (\ie LFRR). 
It indicates that increased granularity in interactions has a positive impact on the accuracy of recommendations, even in the case of static representations.
As for sequential recommendation methods, it can be observed that they generally achieve better performance compared to conventional collaborative filtering methods. We suppose this can be attributed to their ability to dynamically model user preferences based on the behavior sequences. 
Furthermore, FMLP-Rec attains considerable results compared to transformer-based models (\ie SASRec, SSE-PT, BERT4Rec). One possible reason for this is that the proposed learnable filtering layer has the ability to efficiently decrease noisy signals.

Additionally, in the recruitment scenario, there are also results of person-job fit methods. We can observe that models that rely exclusively on user resumes and job descriptions (\ie PJFNN and BPJFNN) are frequently subject to severe limitations caused by the quality and volume of textual data. 
These models perform better on Technology datasets that contain higher quality and larger volumes of text data compared to that Design and Sale datasets.
Among the hybrid models that combine both text and collaborative behavior signals (\ie IPJF, PJFFF, DPGNN), DPGNN achieves the best performance, which is inseparable from its modeling capacity of the two-way selection process in reciprocal recommendation.

Overall, compared with all baseline methods, it can be seen that our approach ReSeq achieves the best results in almost all the cases across five datasets from two scenarios. 
Different from these baselines, we formulate reciprocal recommendation as a distinctive sequence matching task, where matching prediction is performed based on the dynamic behavior sequences of both parties. 
And for lower time consumption, we introduce the self-distillation techniques to distill knowledge from the micro to macro level matching, which greatly speeds up the similarity calculation. As a result, our ReSeq is an effective and efficient reciprocal recommendation method.

\subsection{Ablation Study}
The major contribution of our method ReSeq lies in four parts: decomposed embedding, behavior sequence encoding, matching prediction, and efficiency improvement. 
In order to verify the effectiveness of specific designs for each component, we conduct an ablation study on Design and StackOverflow datasets to verify their contributions. 
Specifically, we consider the following four variants of ReSeq:
\begin{itemize}
\item {\underline{ReSeq w/o DSE}} removes the decomposition and sharing operations of embedding matrices (Eqn.~\eqref{eq:decomp}) and directly learns four embedding matrices for users.
\item {\underline{ReSeq w/o MASK}} 
replaces the different mask matrices for encoding user behavior sequences in Section~\ref{sec:seq_encode} with the same bidirectional mask matrix.
\item {\underline{ReSeq w/o TSA}} exchanges the time-sensitive attention aggregation operation in micro-level matching (Eqn.~\eqref{eq:agg}) into mean aggregation.
\item {\underline{ReSeq w/o SD}}  removes the self-distillation loss.
\end{itemize}

From the results in Table~\ref{tab:ablation_study}, it is evident that removing any of the above components would result in a decline in the overall effect. This indicates that all of the components in ReSeq play a vital role in delivering precise recommendations.

\subsection{Further Analysis}
\subsubsection{Hyperparameter Tuning}
In this section, we assess the robustness of our model and conduct a comprehensive analysis of two crucial loss function coefficients. The performance of the two datasets is presented in Figure~\ref{fig:parameters}. 
For each loss coefficient, we show the NDCG@5 of the two perspectives separately and calculate the average metrics of the two perspectives to reflect the overall trend.

For the micro teacher loss coefficient $\lambda$, we tune it within the range from 1 to 10 and additionally designate it as 0.5 to investigate the ramifications of inadequate teacher preparation. 
As shown in Figure~\ref{fig:parameters}, with the increasing of coefficient $\lambda$, the model's performance gradually improves. At the value of 5, ReSeq achieves the best performance, and further increment of $\lambda$ does not result in significant performance degradation, which fully reflects our method's robustness.

For the self-distillation loss coefficient $\mu$, we tune it in the ranges of \{0.0001, 0.001, 0.003, 0.005, 0.01, 0.05, 0.1\}. 
As shown in Figure~\ref{fig:parameters}, our model achieves the best results on the two datasets when $\mu$ is set to 0.005 and 0.01, respectively. A self-distillation loss coefficient that is too small cannot effectively transfer knowledge between micro and macro levels, while a too-large loss coefficient may cause the model to disregard ranking optimization.

\subsubsection{Deployment Efficiency Analysis}
To investigate the impact of micro-to-macro self-distillation on time latency in real-world deployment, we compare the efficiency of matching prediction between SASRec, ReSeq, and ReSeq w/o Self-Distillation on two datasets from different scenarios, \ie Design and StackOverflow.
All experiments are carried out on a Linux system machine with an Intel(R) Xeon(R) CPU E5-2630 v4 @ 2.20GHz CPU and a 12GB NVIDIA TITAN V GPU. For a fair comparison, we use the PyTorch framework to implement all methods and ensure that the common parameters such as batch size are uniform.
Specifically, we count the average matching prediction time of each batch of the three methods on the test sets. Besides, we maintain process synchronization through the \texttt{synchronize} function in PyTorch to obtain accurate time latency.

\begin{figure}[]
\centering
\includegraphics[width=1.0\linewidth]{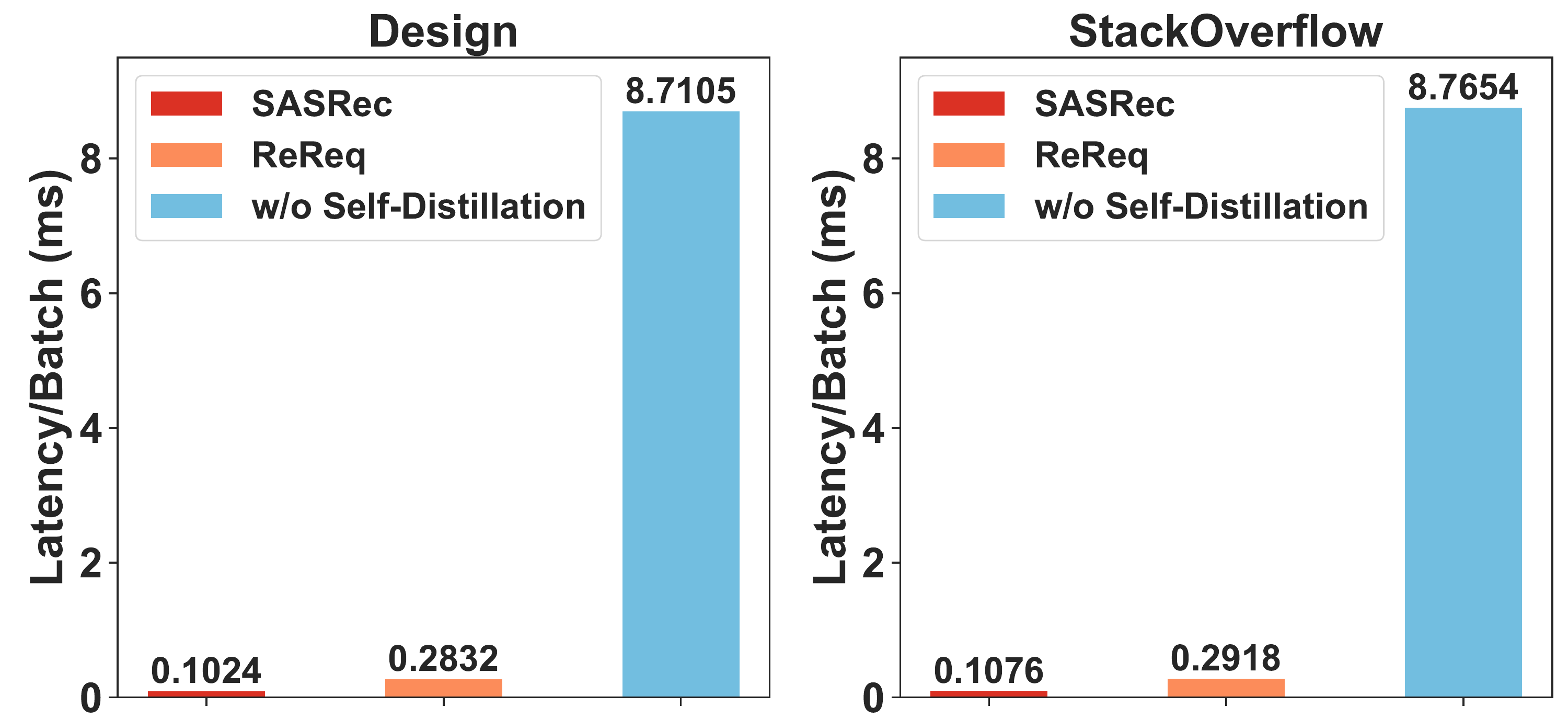}
\caption{Per-batch matching prediction latency.}
\label{fig:efficiency}
\end{figure}

As shown in Figure~\ref{fig:efficiency}, our ReSeq has a latency slightly higher than SASRec. The additional consumption is an inevitable time increase brought by dual-perspective matching in reciprocal recommendation to achieve effective sequence matching.
However, without self-distillation, the fine-grained micro-level matching prediction would result in decades of increase in time consumption, which is unacceptable in actual deployment. In contrast, the micro-to-macro self-distillation enables our ReSeq to be an efficient and effective reciprocal recommendation method.

\begin{figure}[]
\centering
\includegraphics[width=1.0\linewidth]{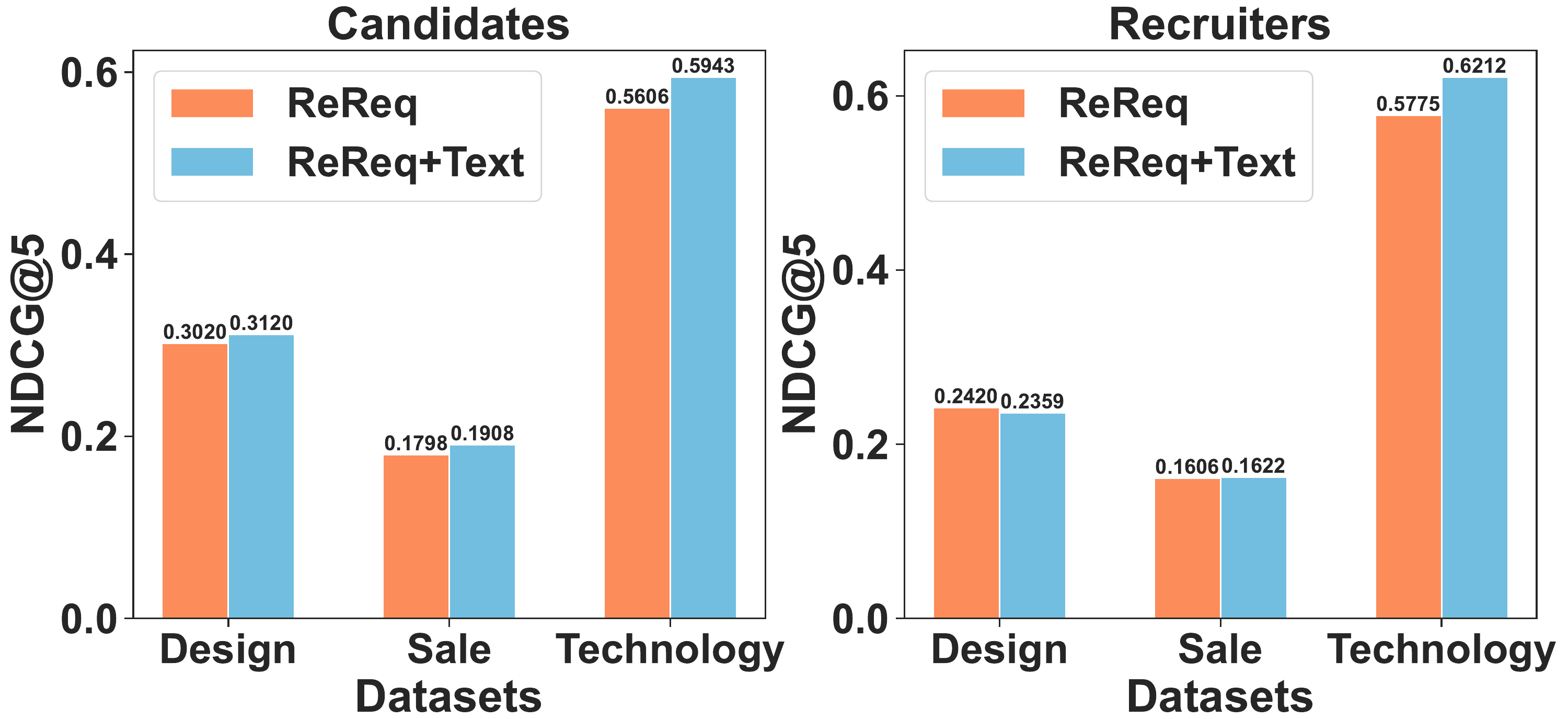}
\caption{Performance after adding text-based matching.}
\label{fig:addtext}
\end{figure}

\subsubsection{Text Enhancement}
Out of all, we conduct a text enhancement experiment on recruitment datasets. In detail, we directly incorporate a text matching score derived from pre-trained twin-tower BERT with an additional prediction layer into the macro-level matching prediction (Eqn.~\eqref{eq:ma_score}). The results are presented in Figure~\ref{fig:addtext}. Observably, after combining with simple text enhancement, our method basically obtains performance improvements in both perspectives on the three datasets.
Particularly, ReSeq benefits more on the Technology dataset.
The potential explanation for this phenomenon is that the text quality and quantity vary prominently across different datasets. Thus, in the Technology dataset, whose text quality and quantity substantially surpass the other two datasets, the effect of text enhancement is more significant. 
The above results illustrate that our method can be improved by incorporating some supplementary information.
Certainly, this simple text enhancement method may impose some limitations, and in the future, we will take into account more rational approaches to merge text or other information.

\section{Conclusion}
This paper proposes a reciprocal sequential recommendation method, named \emph{ReSeq}. To model dynamic user interests from both sides, we formulate reciprocal recommendation as a distinctive sequence matching task.
Specifically, we encode user behavior sequences from both active and passive perspectives with different module designs.
Subsequently, we conduct comprehensive and fine-grained sequence interactions at both macro and micro levels through multi-scale sequence matching.
In addition, to alleviate the increased time consumption caused by fine-grained interactions, we improve the model deployment efficiency via micro-to-macro self-distillation.
Extensive experiments on real-world datasets show that our approach outperforms several competitive baselines from two perspectives in RRS.
In the future, 
we will consider incorporating multiple auxiliary sequential features (\eg textual information and categories) into our framework in a universal, effective, and efficient way. 
Besides, we would like to further alleviate the sparsity issue while training reciprocal sequential models, including pre-training, data augmentation, and self-supervised learning techniques.

\balance

\bibliographystyle{ACM-Reference-Format}
\bibliography{ref}

\end{document}